\documentstyle[pre,aps]{revtex}

\newcommand{\be}{\begin{equation}}
\newcommand{\ee}{\end{equation}}

\begin{document}

\draft

\title{Booms and Crashes in Self--Similar Markets} 

\author{S. Gluzman$^1$ and V. I. Yukalov$^2$\footnote{The author to 
whom correspondence is to be addressed}} 

\address{$^1$International Center of Condensed Matter Physics\\
University of Brasilia, CP 04513, Brasilia, DF 70919-970, Brazil \\
and \\
$^2$Centre for Interdisciplinary Studies in Chemical Physics\\
University of Western Ontario, London, Ontario N6A 3K7, Canada}

\maketitle

\vspace{3cm}

\begin{abstract}

Sharp changes in time series representing market dynamics are studied by
means of the self--similar analysis suggested earlier by the authors.
These sharp changes are market booms and crashes. Such crises phenomena in
markets are analogous to critical phenomena in physics. A simple
classification of the market crisis phenomena is given.

\end{abstract}

\newpage

\section{Introduction}

Booms and crashes of market structures are examples of dynamical transitions 
that remind phase transitions and critical phenomena in physical systems. 
This analogy is, of course, not direct, since markets are nonequilibrium 
systems. Nevertheless, it is possible to develop a renormalization group 
approach for analysing the dynamics of markets near the points of their 
critical changes, such as booms and crashes, somewhat analogous to the 
renormalization group technique used for describing critical phenomena. Since
nonequilibrium markets are quite different from equilibrium systems of 
statistical mechanics, the renormalization group needed for the analysis of
market behaviour is to be also very different from the statistical 
renormalization group. We show that such a natural tool for describing 
market critical phenomena is the algebraically invariant self--similar
renormalization group. In this approach, a market critical restructuring
is treated analogously to a critical phenomenon in physics, which permits us
to present the market time evolution in the critical region by means of a 
group property called, in physical language, the property of self--similarity. 
Using this self--similar renormalization group allows us to renormalize 
the time series generated by the market and to describe the occurrence of 
booms and crashes in good quantitative agreement with the data available.

A powerful tool for describing critical phenomena in equilibrium statistical 
systems is the so--called statistical renormalization group, formulated 
by Wilson being based on the Kadanoff idea of scaling transformations. A
detailed account of this technique can be found in many books, for instance, 
in Ma [1]. Markets can be considered as very complex nonequilibrium
statistical systems, and sharp changes of market structure, such as 
crashes, can be thought of as a kind of nonequilibrium phenomena, comparable
to earthquakes [2,3]. There have been suggested several other analogies 
between physical systems and markets [2-5]. However, to our knowledge, 
there has been no attempt to develop a renormalization group approach for 
treating the behaviour of markets near their critical points of sharp 
changes, that is near their booms and crashes. We have suggested such an
approach in our previous papers [6,7].

As far as markets, despite some analogies with physical systems, are much
more complex and, in addition, nonequilibrium, the standard statistical
renormalization group cannot be applied to such transitions as booms and 
crashes. Fortunately, there exists another approach based on self--similar
renormalization group, which provides an evolution equation in the space 
of approximations [8-12]. This approach, completed by the condition of 
algebraic invariance, makes the basis of the algebraic self--similar 
renormalization [13-15]. Dynamics in the space of approximations,
for the case of a market, is nothing but the time evolution of this market. 
This is why the self--similar renormalization group seems to be an absolutely 
natural concept for treating market evolution.

Let us stress the main ideas explaining why our approach suites well for 
describing market dynamics: (i) Sharp structural changes in a market are
equivalent to critical phenomena in a physical system. (ii) The evolution 
of a market in the critical region can be formulated by means of a group 
property with respect to time, similarly to the evolution of a Hamiltonian
with respect to its parameters in statistical renormalization group. 
(iii) In the same way as in the latter group, where in order to get a 
good quantitative description, one needs to make just a few renormalization 
steps, in the self--similar renormalization group, we also need to take into 
account only a few temporal points.

\section{Scheme of Analysis}

Here we present a short survey of the self--similar analysis to be used in
what follows. More details can be found in our previous papers [6,7]. The
main characteristic of any market activity is the price for a security or
commodity, or some index related to this price. Let $f(t)$ be such a
characteristic of a market at the moment of time $t$. A sharp change of
the value of price, that is of $f(t)$, occurring during the period of time
comparable with the resolution of the time series corresponding to $f(t)$,
is commonly called a crisis. Booms are sharp upward moves and crashes are
sharp downward moves of the price.

Suppose that the values of $f(t)$ are known for $n+1$ time points
$t=k=0,1,2,\ldots,n$, so that
\be
f(k) = a_k \qquad (k=0,1,\ldots,n).
\ee
The aim of our analysis is to predict, being based on the data (1), the
behaviour of $f(t)$ for $t\geq n+1$.

At the first step, for the function $f(t)$ in the time interval $0\leq
t\leq n$, we construct a polynomial representation
\be
p_n(t) =\sum_{k=0}^n A_kt^k \qquad (0\leq t\leq n) ,
\ee
in which the coefficients $A_k$ are defined by the equations
\be
p_k(k)=a_k \qquad (k=0,1,\ldots,n) .
\ee
The coefficients of polynomial (2) characterize different tendencies, or
trends, existing in the market. The plus or minus sign of $A_k$ describes
the tendency to growth or to decrease, respectively. Such different trends
may be called {\it heterotrends}. The latter are somewhat analogous to
heterophase fluctuations in statistical systems [16], where they play a
very important role, especially in the vicinity of phase transitions.
Market models must also include some kind of heterogeneity in their 
structure [17,18].

The competition of different tendencies in a market determines the
formation of following prices. If there are no constraints imposed from
outside, the market should develop according to its own laws. We assume
that this natural dynamics of a market can be formulated as the property
of self--similarity for a given time series.

The algebraic self--similar renormalization group selects the most stable
nonlinear mixture of tendencies prevailing over the less stable. Thus,
evolving and competing via the self--similar renormalization dynamics,
different trends form a {\it self--similar heterotrend market}, or simply
{\it self--similar market}. The problem of forecasting the price for a
real market is mapped to the problem of evolution of a self--similar
market. Let us emphasize again that the  possibility of so simplifying
market dynamics is justified in the critical region where collective 
coherent effects become prevailing. The development of such a coherent
behaviour with strong correlations between market agents is a necessary
condition for the formation of a law of collective motion, which, in turn, 
can be expressed through the self--similar renormalization group.

For the sequence $\{ p_k(t)\}$ of polynomials $p_k(t)=\sum_{m=0}^k
A_mt^m$, we may define [15] the self--similar exponential approximants
\be
f_k(t,\tau) =A_0\exp\left (\frac{A_1}{A_0}t\exp\left (\frac{A_2}{A_1}t
\ldots\exp\left (\frac{A_k}{A_{k-1}}\tau t\right )\ldots\right )\right ) ,
\ee
where it is assumed that $A_m\neq 0\; (m=0,1,\ldots,k)$. This kind of nested 
exponentials, because of their self--similar structure, is, as we think, the 
most suitable functional form for describing market dynamics. The effective
time $\tau$ is to be found from a fixed--point condition, e.g., taken as
the minimal--difference condition
\be
\left | f_n(t,\tau_n) - f_{n-1}(t,\tau_n)\right | =\min_{\tau}\left |
f_n(t,\tau) - f_{n-1}(t,\tau)\right | ,
\ee
giving $\tau_n=\tau_n(t)$. In many cases, condition (5) can be reduced to
\be
f_n(t,\tau) = f_{n-1}(t,\tau), \qquad \tau=\tau_n(t) ,
\ee
which, according to (4), is equivalent to the equation
\be
\tau =\exp\left (\frac{A_n}{A_{n-1}}\tau t\right ) , \qquad \tau=\tau_n(t) .
\ee
The latter, as is obvious, possesses a solution if $A_n/A_{n-1}<0$, but
when $A_n/A_{n-1}>0$, solutions may be absent. With the found effective
time $\tau_n(t)$, we obtain the self--similar forecast
\be
f_n^*(t)\equiv f_n(t,\tau_n(t)) \qquad (t\geq n+1) .
\ee

It may happen that, when $A_n/A_{n-1}>0$, equations (6) and (7) have no 
solutions. Then we need to return to condition (5) which, together with
(4), yields $\tau_n=0$. In this case, the forecast becomes
\be
f_n^*(t) =f_{n-1}(t,1) \qquad (\tau=0) .
\ee

The stability analysis of the procedure is checked by analysing the
multipliers
\be
M_k(t,\tau) \equiv
\frac{\delta f_k(t,\tau)}{\delta f_1(t,1)} , \qquad M_k(t)\equiv M_k(t,1) ,
\ee
where $k=1,2,\ldots,n$. Note that $M_1(t,1)=1$ for any $t$. When Eq. (6)
has a solution $\tau_n(t)$, then the multiplier at the fixed point (8) is
defined as
\be
M_n^*(t)\equiv\frac{1}{2}\left [ M_n(t,\tau_n(t)) + M_{n-1}(t,\tau_n(t))
\right ] .
\ee
When (6) has no solution, then the fixed point is given by (9), and the
related multiplier is
\be
M_n^*(t) = M_{n-1}(t,1) \qquad (\tau = 0) .
\ee

It is also admissible to define a fixed point as an average
\be
\stackrel{-}{f}_n(t) \equiv\frac{1}{2}\left [ f_n(t,1) + f_{n-1}(t,1)
\right ] .
\ee
In this case, the corresponding multiplier can be written as
\be
\stackrel{-}{M}_n(t) \equiv\frac{1}{2}\left [ M_n(t,1) + M_{n-1}(t,1)
\right ] .
\ee

The forecasting procedure is stable if the multiplier at the fixed point
satisfies the inequality $|M_m^*(t)|\leq 1$ or, respectively,
$|\stackrel{-}{M}_n(t)|\leq 1$. The case of an equality is called
neutrally stable. As an optimal forecast, that corresponding to the
minimal multiplier is to be taken.

Note that the difference
$$ \Delta_n(t) \equiv f_n(t,1) - f_{n-1}(t,1) $$
describes the degree of a market volatility. For a very volatile market,
$\Delta_n(t)$ can be comparable with $f_n(t,1)$. Contrary to this, for a
steady--state market, $\Delta_n(t)$ is much less than $f_n(t,1)$.

In the following section, we pass to the consideration of particular
examples of self--similar markets. All data, unless stated otherwise, are
taken from the books of International Financial Statistics issued by the
International Monetary Fund and from UNCTAD Commodity Yearbooks issued by
the United Nations.

We start the self--similar analysis from the simplest case, when the
values of $f(t)$ are given for only three time points. So, we shall deal
with the self--similar exponential approximants
$$ f_1(t,\tau)=a_0\exp\left (\frac{A_1}{a_0}\tau t\right ) , \qquad
f_2(t,\tau) =a_0\exp\left (\frac{A_1}{a_0}t\exp\left (\frac{A_2}{A_1}
\tau t\right )\right ) $$
and the multipliers
$$ M_1(t,\tau)=\tau\exp\left\{ -\frac{A_1}{a_0}\left ( 1 -\tau \right ) t
\right\} , $$
$$ M_2(t,\tau) =\left ( 1 +\frac{A_2}{A_1}\tau t\right )\exp\left\{
-\frac{A_1}{a_0}\left [ 1 -\exp\left (\frac{A_2}{A_1}\tau t\right )
\right ] t + \frac{A_2}{A_1}\tau t\right \} . $$
It turns out that specific features of a market are directly related to
the quantities
\be
X \equiv a_1 - a_0 , \qquad Y \equiv a_2 - a_1 .
\ee
The coefficients of the polynomial representation (2)  can be expressed
through these quantities as
$$ A_0 = a_0, \qquad A_1 =\frac{1}{2}\left ( 3X - Y\right ) , \qquad
A_2 =\frac{1}{2}\left ( Y - X\right ) . $$
Depending on the values of $X$ and $Y$, different possibilities can arise.

\section{Upward Bound Markets}

This case implies that $a_0\leq a_1< a_2$, hence $X\geq 0,\; Y>0$. The
following subcases exist.

\subsection{Balanced Bull Market ($0<Y<X$)}

Then $A_1>0,\; A_2<0$, and $|A_1|>|A_2|$. The bullish tendency presented
by $A_1$ dominates over the bearish tendency given by $A_2$. For the
balanced bull market, we have $|M_2(3)|<1$, and prices rise. When a price
goes up sharply, we get a boom. Some examples of such booms are considered
below.

\vspace{3mm}

(i) The average index of the Swedish share prices ({\sl 1980}=100) from
{\sl 1980} to {\sl 1982}:
$$ a_0=100\;({\sl 1980}), \qquad a_1=149\; ({\sl 1981}), \qquad
a_2=185\; ({\sl 1982}). $$
Let us find the index for {\sl 1983}. The coefficients of polynomial (2),
defined by (3), are $A_1=55.5$ and $A_2=-6.5$. For the exponential
approximants (4), we have $f_1(3,1)=528.567$ and $f_2(3,1)=322.754$.
Recall that $M_1(t)\equiv 1$ for any $t$, and using (10), we get
$M_2(3)=0.279$. For the average (13), we find $\stackrel{-}{f}_2(3)=425.661$,
with the multiplier (14) being $\stackrel{-}{M}_2(3)=0.639$. From the
minimal--difference condition (7), it follows that $\tau=0.76445$. This
gives for approximant (8) the value $f_2^*(3)=357.087$, and for multiplier
(11), we have $|M_2^*(3)|=0.447$. Since $M_2^*(3)$ has the absolute value
which is less than that of $\stackrel{-}{M}_2(3)$, we have to accept
$f_2^*(3)$ as the optimal forecast. The actual index boomed in {\sl 1983}
to $359$. Thus, the error of our forecast is $-0.5\%$.

\vspace{3mm}

(ii) The average index of the Philippines share prices (${\sl 1985}=100$)
from the fourth quarter of {\sl 1986} till the second quarter of {\sl
1987}:
$$ a_0=224.9\;(IV,\;{\sl 1986}), \qquad a_1=273\; (I,\; {\sl 1987}),
\qquad a_2=315.6\; (II,\;{\sl 1987}) . $$
What is the index in the third quarter of {\sl 1987}? Following the
standard prescription, we find the polynomial coefficients $A_1=50.85$ and
$A_2=-2.75$. Then we get $f_1(3,1)=443.172$ and $f_2(3,1)=400.363$, with
the multiplier $M_2(3)=0.643$. From here, $\stackrel{-}{f}_2(3)=421.767$
with $\stackrel{-}{M}_2(3)=0.842$. The minimal--difference condition gives
$\tau=0.8686$, so that $f_2^*(3)=405.371$ with $|M_2^*(3)|=0.739$. The
actual value of the index in the third quarter of {\sl 1987} was $424.53$.
Both our estimates are pretty close to this value, with an error less than 
$4.5\%$.

\subsection{Super--Bull Market ($0\leq X < Y <3X$)}

Only bullish tendencies are present, since $A_1>0$ and $A_2>0$. For this
case, Eq. (7) may have no solution, then the minimal--difference condition
(5) yields $\tau=0$. Therefore, instead of (8), we must consider (9).
Exponential approximants increase with time, and $|M_2(3)|>1$. An
interesting situation may occur -- Despite an exponential growth of the
approximants, the value of the sought function at $t=n+1$ can be less than
$a_n$. And even a crash may happen. We illustrate this by the examples
below.

\vspace{3mm}

(i) The average index of the Canadian industrial share prices 
(${\sl 1990}=100$) from {\sl 1967} to {\sl 1969}:
$$ a_0=25.9\; ({\sl 1967}) , \qquad a_1=27.2\; ({\sl 1968}), \qquad
a_2=30.3\; ({\sl 1969}) . $$
Let us look for the index in {\sl 1970}. Again following the prescribed
procedure, we get the polynomial coefficients $A_1=0.4$ and $A_2=0.9$. The
exponential approximants are $f_1(3,1)=27.128$ and $f_2(3,1)\sim 10^{18}$.
The latter is not probable because of $M_2(3)$ being of similar order. The
same holds true for $\stackrel{-}{f}_2(3)$ for which
$|\stackrel{-}{M}_2(3)|\gg 1$. Consequently, the optimal forecast is
$f_2^*(3)=f_1(3,1)=27.128$, with $M_2^*(3)=1$. The actual index in {\sl
1970} was down to $26.6$. Our prediction gives an error of $2\%$.

\vspace{3mm}

(ii) The Malaysian gross domestic product (${\sl 1985}=100$) from {\sl
1987} to {\sl 1989}:
$$ a_0=82.585\;({\sl 1987}), \qquad a_1=89.967\; ({\sl 1988}) , \qquad
a_2=97.804\; ({\sl 1989}) . $$
What would be the value of GDP in {\sl 1990}? The polynomial coefficients
are $A_1=7.154$ and $A_2=0.227$. Then $f_1(3,1)=107.096$ and 
$f_2(3,1)=109.918$, with $M_2(3)=1.237$. Thence,
$\stackrel{-}{f}_2(3)=108.507$ with $\stackrel{-}{M}_2(3)=1.119$, while
$\tau=0$ and $f_2^*(3)=f_1(3,1)=107.096$, with $M_2^*(3)=1$. Consequently,
the optimal forecast is $f_2^*(3)$, which deviates from the actual value
$107.406$ in {\sl 1990} by only $0.29\%$.

\vspace{3mm}

(iii) The market tea prices (in US dollars per metric ton) at the Average
Auction, London, in {\sl 1982--1985}:
$$ a_0=1931.7\; ({\sl 1982}) , \qquad a_1=2324.6\; ({\sl 1983}), \qquad
a_2=3456.8\; ({\sl 1984}) . $$
Let us find the price in {\sl 1985}. The polynomial coefficients are
$A_1=23.25$ and $A_2=369.65$. Since for the approximant
$\stackrel{-}{f}_2(3)$, we have $|\stackrel{-}{M}_2(3)\gg 1$, the optimal
forecast is $f_2^*(3)=f_1(3,1)=2003$, with $M_2^*(3)=1$. The actual price
in {\sl 1985} was down  to $1983.6$. The error of our forecast is $0.98\%$.

\subsection{Bull--Turned--to--Bear Market ($0\leq 3X < Y$)}

This implies $A_1<0,\; A_2>0$, and $|A_1|<|A_2|$. The bullish tendency
presented by $A_2>0$ is not strong enough to prevent the price from
falling down, because of the influence of the bearish tendency given by
$A_1<0$. The growth of prices is too fast, forming a bubble that can
burst. Examples of such bubble crashes are given below.

\vspace{3mm}

(i) The average index of the Swiss industrial share prices 
(${\sl 1980}=100$) from the first quarter of {\sl 1987} to the third
quarter of {\sl 1987}:
$$ a_0=206.9\; (I,\; {\sl 1987}), \qquad a_1=211.5\; (II,\; {\sl 1987}) ,
\qquad a_2=245.5\; (III,\; {\sl 1987}) . $$
What is the index in the fourth quarter of {\sl 1987}? For the polynomial
coefficients we have $A_1=-10.1$ and $A_2=14.7$. The approximants are
$f_1(3,1)=178.714$ and $f_2(3,1)=206.516$, with $M_2(3)=-0.049$. The
average is $\stackrel{-}{f}_2(3)=192.615$, with
$\stackrel{-}{M}_2(3)=0.476$. From the minimal difference condition (7),
we get $\tau=0.28638$. Then, $f_2^*(3)=198.402$, with $|M_2^*(3)|=0.119$.
The optimal forecast is $f_2^*(3)$. The actual index in the fourth quarter
of {\sl 1987} was $202.5$. The error of our forecast is $-2.024\%$.

\vspace{3mm}

(ii) The tin prices (in US cents per pound) in Bolivia from {\sl 1987} to
{\sl 1989}:
$$ a_0=308.97\; ({\sl 1987}) , \qquad a_1=320.22\; ({\sl 1988}) , \qquad
a_2=400.71 \; ({\sl 1989}) . $$
Let us find the price in {\sl 1990}. The polynomial coefficients are
$A_1=-23.37$ and $A_2=34.62$. For the approximants we have
$f_1(3,1)=246.246$ and $f_2(3,1)=308.148$, with $M_2(3)=-0.051$. Hence,
$\stackrel{-}{f}_2(3)=277.197$, with $\stackrel{-}{M}_2(3)=0.475$. The
minimal difference condition gives an effective time $\tau=0.2836$. Then,
$f_2^*(3)=289.714$, with $|M_2^*(3)|=0.124$. The optimal forecast is
$f_2^*(3)$, deviating from the actual price $286.88$ in {\sl 1990} by
$0.99\%$.

\vspace{3mm}

(iii) The volume of the Argentine export (in billions of US dollars) in
{\sl 1979--1981}:
$$ a_0=7.81\; ({\sl 1979}), \qquad a_1=8.021\; ({\sl 1980}), \qquad
a_2=9.143\; ({\sl 1981}). $$
Let us look for the volume in {\sl 1982}. The coefficients are
$A_1=-0.244$ and $A_2=0.455$. Then, $f_1(3,1)=7.110$ and $f_2(3,1)=7.807$,
with $M_2(3)=-0.019$. From here, $\stackrel{-}{f}_2(3)=7.459$, with 
$\stackrel{-}{M}_2(3)=0.491$. The effective time is $\tau=0.2489$, which
yields $f_2^*(3)=7.63$, with $|M_2^*(3)|=0.082$. The optimal forecast is
$f_2^*(3)$. The actual volume in {\sl 1982} was $7.625$. The error of our
forecast is $0.066\%$.

\section{Downward Bound Markets}

In this case $a_0\geq a_1 > a_2$, because of which $X\leq 0$ and $Y<0$.
Three possibilities can arise.

\subsection{Balanced Bear Market ($X<Y<0$)}

This means that $A_1<0,\; A_2>0$, and $|A_1|>|A_2|$. The bearish tendency
in the negative $A_1$ dominates over the bullish tendency in the positive
$A_2$, which leads to the decrease of price. Examples are given below.

\vspace{3mm}

(i) The exchange rate of Japanese Yen to SDR (Special Drawing Rights) in
{\sl 1984--1986}:
$$a_0=246.13\; ({\sl 1984}), \qquad a_1=220.23\; ({\sl 1985}), \qquad
a_2=194.61\; ({\sl 1986}) . $$
Let us find the rate in {\sl 1987}. The coefficients are $A_1=-26.04$ and
$A_2=0.14$. Repeating the same steps as above, we have
$f_1(3,1)=179.194,\; f_2(3,1)=180.106$, with $M_2(3)=0.973$, which yields
$\stackrel{-}{f}_2(3)=179.65$ with $\stackrel{-}{M}_2(3)=0.986$. The
effective time is $\tau=0.98425$. Then $f_2^*(3)=180.092$ with
$|M_2^*(3)|=0.981$. Both forecasts are very close to each other as well as
to the actual rate $175.2$ in {\sl 1987}, the error being less than $3\%$.

\vspace{3mm}

(ii) The unit value of the Argentine wheat exports (${\sl 1990}=100$) in
{\sl 1982--1984}:
$$ a_0=115.6\; ({\sl 1982}), \qquad a_1=99.4\; ({\sl 1983}), \qquad
a_2=91.6\; ({\sl 1984}). $$
What would be the value in {\sl 1985}? In the standard way, we get
$A_1=-20.4,\; A_2=4,2,\; f_1(3,1)=68.083,\; f_2(3,1)=86.892,\; M_2(3)=0.263$,
so that $\stackrel{-}{f}_2(3)=77.488,\;\stackrel{-}{M}_2(3)=0.631$. Then
$\tau=0.663696$, and $f_2^*(3)=81.351,\; |M_2^*(3)|=0.631$. Again, both
forecasts are equivalent from the viewpoint of stability, close to each
other and to the actual value $79.8$ in {\sl 1985}, the error being less
than $3\%$.

\vspace{3mm}

(iii) The average index of the Korean share prices (${\sl 1985}=100$) in
{\sl 1989--1991}:
$$ a_0=661.2\; ({\sl 1989}), \qquad a_1=537.7\; ({\sl 1990}), \qquad
a_2=473\; ({\sl 1991}) . $$
Let us find the index for {\sl 1992} and compare it with that $422.63$
actually happened. In the usual way, we get $A_1=-152.9,\; A_2=29.4,\;
f_1(3,1)=330.4,\; f_2(3,1)=447.83,\; M_2(3)=0.322$, and
$\stackrel{-}{f}_2(3)=389.116$, with $\stackrel{-}{M}_2(3)=0.661$. The
effective time is $\tau=0.6768$, which results in $f_2^*(3)=413.453$, with
$|M_2^*(3)|=0.681$. The error of $\stackrel{-}{f}_2(3)$ is $-7.9\%$ and
that of $f_2^*(3)$ is $-2.2\%$.

\subsection{Super--Bear Market ($3X<Y<X\leq 0$)}

This implies that both $A_1$ and $A_2$ are negative. Only bearish
tendencies are presented. As an example, consider the volume of GDP (in
millions of SDR) of the Yemen Republic in {\sl 1979--1981}:
$$ a_0=1084\; ({\sl 1979}), \qquad a_1=1006\; ({\sl 1980}), \qquad
a_2=826\; ({\sl 1981}). $$
Let us calculate the volume in {\sl 1982}, comparing it with the actual
one equal to $502$. We have $A_1=-27,\; A_2=-51,\; f_1(3,1)=1006,\;
f_2(3,1)=4.5\times 10^{-7},\; M_2(3)=8.6\times 10^{-7}$, and
$\stackrel{-}{f}_2(3)=502.976$, with $\stackrel{-}{M}_2(3)=0.5$. The
minimal difference condition (5) gives $\tau=0$. So, $f_2^*(3)=f_1(3,1)=1006$,
with $M_2^*(3)=M_1(3)=1$. The optimal forecast is $\stackrel{-}{f}_2(3)$,
whose error is $0.19\%$.

\subsection{Bear--Turned--to--Bull Market ($Y<3X\leq 0$)}

Now $A_1>0,\; A_2<0$, and $|A_1|<|A_2|$. The bullish trend in positive
$A_1$ becomes dominating, although the bearish trend in negative $A_2$
still remains. The motion looks like a downward bubble that bursts.
Consider some examples.

\vspace{3mm}

(i) The average index of the US industrial share prices (${\sl 1985}=100$)
in {\sl 1968--1970}:
$$ a_0=51.7\; ({\sl 1968}), \qquad a_1=51.2\; ({\sl 1969}), \qquad
a_2=43.9\; ({\sl 1970}) . $$
We shall find the index for {\sl 1971} and compare it with the actual
value $52.1$. We get $A_1=2.9,\; A_2=-3.4,\; f_1(3,1)=61.175,\;
f_2(3,1)=51.96,\; M_2(3)=-0.063$, and $\stackrel{-}{f}_2(3)=56.568$,
with $\stackrel{-}{M}_2(3)=0.469$. The minimal--difference condition gives
$\tau=0.3221$, because of which $f_2^*(3)=54.58$ and $|M_2^*(3)|=0.124$.
The optimal forecast is $f_2^*(3)$, with an error $4.8\%$.

\vspace{3mm}

(ii) The volume of the US imports (${\sl 1985}=100$) in {\sl 1973--1975}:
$$ a_0=57.5\; ({\sl 1973}), \qquad a_1=56.7\; ({\sl 1974}), \qquad
a_2=49.9\; ({\sl 1975}) . $$
Let us find the volume in {\sl 1976}, comparing it with the actual value
$60.8$. We may mention that the decline in imports in these years could be
related to the oil embargo and the energy crisis of {\sl 1973}. Following
the standard procedure, we have $A_1=2.2,\; A_2=-3,\; f_1(3,1)=64.494,\;
f_2(3,1)=57.61,\; M_2(3)=-0.046,\; \stackrel{-}{f}_2(3)=61.052$, and
$\stackrel{-}{M}_2(3)=0.477$. The effective time is $\tau=0.2969$. Then,
$f_2^*(3)=59.493$, with $|M_2^*(3)|=0.108$. The optimal forecast is
$f_2^*(3)$. with an error $-2\%$, though $\stackrel{-}{f}_2(3)$ is also
good.

\vspace{3mm}

(iii) The ratio of the total Non--Gold Reserves to Imports for Industrial 
Countries in {\sl 1967--1969}:
$$ a_0=8.1\; ({\sl 1967}), \qquad a_1=8.1\; ({\sl 1968}), \qquad
a_2=6.8\; ({\sl 1969}) . $$
We make a prediction for {\sl 1970} and compare it with the actual value
$8.9$. As usual, we find $A_1=0.65,\; A_2=-0.65,\; f_1(3,1)=10.305,\;
f_2(3,1)=8.198,\; M_2(3)=-0.079,\;\stackrel{-}{f}_2(3)=9.252$, and 
$\stackrel{-}{M}_2(3)=0.461$. Also, $\tau=0.35,\; f_2^*(3)=8.812$, with
$|M_2^*(3)|=0.142$. The forecast $f_2^*(3)$ is optimal, with an error
$-0.989\%$.

\section{Yo--Yo Markets}

In such markets, time data are not consecutively ordered but behave
oscillatory. With only three time points we consider here, there can be
two possibilities.

\subsection{Yo--Yo Decline Market ($Y\leq 0 < X$)}

In this case $a_0 < a_1,\; a_1 \geq a_2$, hence $A_1 >0,\; A_2 <0$, and
$|A_1|>|A_2$. The bullish trend in positive $A_1$ overweights the bearish
trend in negative $A_2$. The price, after $a_2$, can go only up. This
behaviour may be called {\it yo--yo boom}. Consider an example.

\vspace{2mm}

The average index of the UK share prices (${\sl 1958}=100$) in {\sl
1960--1962}:
$$ a_0=166\; ({\sl 1960}), \qquad a_1=171\; ({\sl 1961}), \qquad
a_2=158\; ({\sl 1962}) . $$
Let us estimate the index for {\sl 1963} and compare it with the actual
value $181$. Calculations give us $A_1=14,\; A_2=-9,\; f_1(3,1)=213.791,\;
f_2(3,1)=172.219,\; M_2(3)=-0.109$, and $\stackrel{-}{f}_2(3)=193$, with
$\stackrel{-}{M}_2(3)=0.446$. The effective time is $\tau=0.433459$,
thence $f_2^*(3)=185.241$, with $|M_2^*(3)|=0.215$. The optimal forecast
$f_2^*(3)$ has an error $2.3\%$.

\subsection{Yo--Yo Rise Market ($X<0\leq Y$)}

Here we have $a_0>a_1,\; a_1\leq a_2$, so that $A_1<0,\; A_2>0$, and
$|A_1|>|A_2|$. The bearish trend prevails over bullish, as a result, the
price falls below $a_2$. This type of behaviour may be termed {\it yo--yo
crash}. An example is given below.

\vspace{2mm}

The Chilean total reserves (in millions of SDR) in {\sl 1972--1974}:
$$ a_0=204\; ({\sl 1972}) , \qquad a_1=137\; ({\sl 1973}) , \qquad
a_2=149\; ({\sl 1974}) . $$
We make a forecast for {\sl 1975}, comparing it with the actual value of
$84$. Recall that an upheaval happened in {\sl 1974} in Chile. We find
$A_1=-106.5,\; A_2=39.5,\; f_1(3,1)=42.604,\; f_2(3,1)=121.918,\;
M_2(3)=-0.106,\;\stackrel{-}{f}_2(3)=82.261$, and
$\stackrel{-}{M}_2(3)=0.447$. Then $\tau=0.54519$, which leads to
$f_2^*(3)=86.856$, with $|M_2^*(3)|=0.774$. Here the optimal forecast is
$\stackrel{-}{f}_2(3)$ having an error $-2\%$.

\section{Bubble Burst}

The relatively simple three--point classification of market behaviour we
have presented does not exhaust all possible types of market dynamics
which might occur when more points of a time series are taken, but rather
demonstrates how the self--similar analysis can be used for describing
market crises. It is needless to recall that the possibility of predicting
market crises is very much important. We would like now to pay a special
attention to market bubble bursts showing that, in the framework of the
self--similar analysis, all of them occur in the same way. Such an analogy
between bubble bursts of quite different markets makes it possible to
predict these burst phenomena.

\vspace{3mm}

(i) The average index of the Belgian industrial share prices (${\sl
1990}=100$) in {\sl 1971--1973}:
$$ a_0=34\; ({\sl 1971}), \qquad a_1=38\; ({\sl 1972}), \qquad 
a_2=46\; ({\sl 1973}) . $$
Let us look for the index in {\sl 1974} and compare it with the actual
value $37$. Repeating the standard steps, we get $f_1(3,1)=40.562$, while
$f_2(3,1)\sim 10^3$, with $M_2(3)\sim 10^3$. Therefore, the average
estimate $\stackrel{-}{f}_2(3)$ has negligible probability to happen. The
minimal--difference condition (5) gives $\tau=0$. So,
$f_2^*(3)=f_1(3,1)=40.562$, with $M_2^*(3)=1$. The error of $f_2^*(3)$ is
$9.6\%$.

\vspace{3mm}

(ii) The average index of the Japanese share prices (${\sl 1985}=100$) in
{\sl 1987--1989}:
$$ a_0=196.4\; ({\sl 1987}), \qquad a_1=213.9\; ({\sl 1988}) , \qquad
a_2=257.8\; ({\sl 1989}) . $$
During these years, Japan had, what it would have been termed later, the
bubble economy [19]. In {\sl 1990}, the index dropped to $218.8$. Applying
the self--similar analysis to the data above, we find $f_1(3,1)=209.733$,
while $f_2(3,1)$ as well as $M_2(3)$ are again unreasonably large. For the
effective time, we again get $\tau=0$. In this way, the sole stable
estimate is $f_2^*(3)=f_1(3,1)=209.733$, whose error is $-4.1\%$.

\vspace{3mm}

We should keep in mind that the error of three--point forecasts may be of
about $10\%$. For a more accurate analysis, we need to consider a larger
number of historical points. But in the present paper, we wish to limit
ourselves by the simplest, although the crudest, type of the three--point
analysis.

\section{Conclusion}

We have shown how the self--similar analysis can be used for describing
and forecasting various crises. The latter, according to their strength,
can be roughly divided into two types: (1) First degree crash (boom), when
the price falls (grows) by an amount comparable to the gain (loss)
achieved during the period of observation. The majority of bubble crashes
and bubble booms are of this type. Mainly those who buy stocks on margin,
e.g. bears selling short, are hurt. (2) Second degree crash (boom), when
the price loss (gain) is of the order of the price value at the last point
of observation. Yo--yo crashes and booms are often of this category. Then,
not only margin--players can be ruined, but those who own stocks outright
can be hurt too.

We illustrated our approach by several examples. The number of the latter
could be increased to any desirable quantity, since we have considered
hundreds of such cases. In the majority of these cases, the simple
three--point analysis gives reasonable predictions. When the accuracy of
such a simplest analysis is not good enough, one has to invoke additional
information taking into account more data on the market behaviour. However
in this paper, we would like to limit ourselves by the most simple case
involving just three points of data. How to deal with time series
containing more points was briefly explained in Refs. [6,7] and will be
discussed in detail in our following publications.

The principal difference of our approach to describing market crises, as
compared to that based on modelling the market behaviour by a complicated
system of nonlinear differential or difference equations [20], is in the
following. Markets are so complex systems that they can be treated by
dynamical models, with more or less success, only in a stationary state.
But crises are principally nonstationary phenomena. The complexity of real
markets makes it impossible, to our mind, to model them by any complicated
system of nonlinear equations. Moreover, there is quite known property of
nonlinear equations, when adding or omitting a negligibly small term
drastically changes the behaviour of solutions. We do not try to invent,
which, as we think, is impossible, a dynamical model for each market.
Instead, we assume that each market develops self--similarly, according to
its own laws. The result of this development is exhibited in market
characteristics, which contain the hidden information on the laws
governing the market. The self--similar analysis permits one to extract
this hidden information. Thus, each market itself prescribes its future
development -- and this is the essence of the notion of a self--similar
market.

\vspace{5mm}

{\bf Acknowledgement}

\vspace{2mm}

One of us (V.I.Y.) is grateful for financial support to the University of
Western Ontario, Canada.

\end{document}